# Very large spontaneous electric polarization in BiFeO$_3$ single crystals at room temperature and its evolution under cycling fields.

D. Lebeugle, D. Colson, A. Forget, M. Viret

Service de Physique de l'Etat Condensé, DSM/DRECAM, CEA Saclay, 91191 Gif-Sur-Yvette Cedex, France

**Abstract.** Electric polarization loops are measured at room temperature on highly pure BiFeO$_3$ single crystals synthesized by a flux growth method. Because the crystals have a high electrical resistivity, the resulting low leakage currents allow us to measure a large spontaneous polarization reaching 100 µC.cm$^{-2}$, a value never reported in the bulk. During electric cycling, the slow degradation of the material leads to an evolution of the hysteresis curves eventually preventing full saturation of the crystals.

Recently, the perovskite-type oxides which display electric and magnetic orders have raised up renewed interest since the two orders may interact as evidenced by magnetoelectric measurements. There is also a great interest in their potential application in spintronic devices where the magnetic state of some compounds could be affected by an electric field[1]. So far, the vast majority of compounds in which ferroelectricity and magnetism are coupled have low ordering temperatures and room-temperature operation has not been demonstrated yet. $BiFeO_3$ is a good candidate as its space group R3c allows the existence of both antiferromagnetic and ferroelectric orders with very high transition temperatures[2, 3]. $BiFeO_3$ is antiferromagnetic below the Néel temperature $T_N$=643K (neutron powder diffraction[4]) with a long range cycloidal spiral, incommensurate with the lattice[5] ; it is also ferroelectric below $T_C$=1143K (dielectric measurements[6] and X-Ray single crystals diffraction[4]).

From an experimental point of view, revealing the ferroelectric behavior of the multiferroic $BiFeO_3$ at room temperature has proven to be a difficult task. Indeed electrical measurements are very sensitive to the presence of grain boundaries in polycrystalline samples and impurities like inclusions in single crystals. These defects, presumably more conducting than the pure $BiFeO_3$ crystal, induce high leakage currents which generally prevent the application of high electric fields on the sample and make the measurement of a hysteresis loop quite difficult. This remains a persistent problem in the electrical measurements of bulk samples and to date there has been no report of polarization versus electric field loops (P-E loops) at room temperature on single crystals of $BiFeO_3$.

Theoretical studies using density functional theory predict a large ferroelectric polarization of 90-100 $\mu C.cm^{-2}$ [7], consistent with the atomic displacements of the $Bi^{3+}$ and $Fe^{3+}$ ions. The effect is expected not to change significantly between 0K and room temperature because of the high Curie Temperature $T_C$=1143K (extrapolated from electrical measurements made on solid-solutions of $BiFeO_3$ with $Pb(Ti,Zr)O_3$[6]). Recently, a ferroelectric polarization in the range 50 to 90 μC.cm$^{-2}$ has been observed in thin films at room temperature[8]. In ceramic samples, much lower values of the polarization have been measured (8.9 μC.cm$^{-2}$ at room temperature[9,10]) as well as in single crystals (6.1 μC.cm$^{-2}$ at 77K[11]) revealing non saturated hysteretic loops. According to the authors, the difficulty to obtain a well saturated P-E loop comes from the high conductivity of their samples for temperatures above 190K probably due to the presence of impurities in the measured compounds. At the present time, the situation is still unclear concerning the real polarization of the pure compound. There is still a controversy concerning the origin of the large electric polarization found in thin films[8,12]. Indeed, recent calculations show a value of 90-100 μC.cm$^{-2}$ in unstrained $BiFeO_3$[7].

In this letter, we present the polarization loops obtained at room temperature on highly pure single crystals and their evolution under cycling electric fields. The single crystals were grown by a spontaneous crystallization in air from a $Bi_2O_3$-$Fe_2O_3$ flux as detailed in reference 13. With this procedure, we obtain millimeter sized platelets as required for electrical measurements. Electron microprobe analysis confirms the cationic stoichiometry of $BiFeO_3$. The possible oxygen non-stoichiometry was studied by thermogravimetry in controlled atmosphere between 25 and 800°C. No compositional variation was detected in the studied temperature range both under oxygen and argon, confirming a stoichiometric compound. The structural X-ray characterization has been performed on a 0.3x0.2x0.02 mm$^3$ single crystal at 293K with a four-circle Kappa X8 APPEX II Bruker diffractometer (Mo$_{K\alpha}$ radiation, λ = 0.71073 Å). The data of the X-ray analysis revealed a rhomboedrally distorted perovskite-type cell with lattice constants $a_{hex}$ = 5.571 Å and $c_{hex}$ = 13.868 Å. The space group is determined to be R3c with six formula units per unit cell. Three main distortions are responsible for the existence of the spontaneous polarization along the [001]$_{hex}$ direction (i.e. [111]$_{cub}$) : the oxygen octahedra are distorted with minimum and maximum O-O distances of 2.71 and 3.02 Å and rotated by about +/- α =13.8 ° around the threefold axis, the bismuth and iron atoms are shifted by 0.54 and 0.13 Å respectively along the threefold axis[14]. Considering these large atomic displacements, a large electric polarization is expected in the BiFeO$_3$ phase.

The millimeter sized single crystals of BiFeO$_3$ with a high resistivity ($\rho$(300K,100V) ~ 6.10$^{10}$ $\Omega$.cm) have been selected for electrical measurements. A standard 'P(E) loop' method as described in reference 13 has been used. It consists in measuring the current flowing in a simple resistive circuit as a function of the voltage applied to the sample (I(V) characteristic) using a picoamperemeter and voltage source. Silver electrodes are deposited on both sides of the major face of the crystal which corresponds to the (012)$_{hex}$ plane (i.e (110)$_{cub}$). Hence, the perpendicular plane makes an angle of 54°44 with the hexagonal c-axis which contains the easy polarization axis. Measurement of the charge current versus applied electric field during the first electric loading reveals that most of the crystals are single (ferroelectric) domain. Indeed if the electric field is applied in the same direction as the spontaneous polarization, no current flows in the circuit and when the polarity is reversed, we observe an abrupt peak of the current due to the polarization switching of the only electrical domain. When integrating the charge current of two successive switching in the positive and negative voltage, a P(E) cycle can be reconstructed as shown in figure 1 where the first full hysteresis cycle is shown. From this, we extract the polarization and coercive field of our BiFeO$_3$ single crystal at room temperature and we find a remnant polarization P$_{(012)}$ of 60 µC.cm$^{-2}$, a coercive field of 12 kV/cm. Considering our measurement is not along the easy direction (but canted from it by 54°44'), the inferred full saturation polarization along the [001]$_{hex}$ direction is close to 100 µC.cm$^{-2}$. Such a large value of polarization, so far only measured in thin films[8], has been attributed to a structural modification imposed by the substrate. Our measurements on single crystals demonstrate that this polarization is, in fact, an intrinsic property of the BiFeO$_3$ phase as expected theoretically[7].

We have measured several polarization loops in BiFeO$_3$ crystals in order to assess the reproducibility of the cycles and check if any 'fatigue' occurs. Because of their piezoelectric properties, it is well known that electric cycling of ferroelectric materials induces mechanical stresses which degrade the material. As a consequence, the reliability of their electric switching is reduced[15, 16, 17]. This phenomenon is generally called the 'fatigue' of the material and refers to the gradual electrical degradation of the compounds under cycling electric fields above the coercivity.

In the case of BiFeO$_3$ single crystals, we observe a gradual modification of the shape of the P(E) loops during electric field cycling (figure 2). Visually, the loops lose their squareness and the quantities most affected are the coercive and saturation fields as well as the remnant polarization. Observations with an optical microscope confirm that micro-cracks appear during reversal, presumably preventing domain walls from moving easily through the sample. Hence, full saturation is obtained at progressively larger fields for which, eventually, the leakage current induces significant Joule heating. When these large leakage currents are reached, it becomes impossible to fully saturate the crystals and their polarization decreases as can be seen in figure 3. Eventually, temperature effects further damage the samples and the coercive fields cannot be reached anymore. Hence, measurements performed after cycling become unrepresentative of the intrinsic properties of the material.

In summary, we showed here that highly pure single crystals of BiFeO$_3$ have a high enough resistivity to allow us to measure a clear polarization loop at room temperature. The as-synthesized crystals are also found to be ferroelectrically single domain. We infer from the first polarization loop a large intrinsic electric polarization of 100 μC.cm$^{-2}$ at room temperature, which is consistent with that predicted from the perfect BiFeO$_3$ unit cell. Hence, this large value of polarization previously thought to be a property of thin films is in fact intrinsic to the pure BiFeO$_3$ phase. Measurements of several successive polarization loops show that the samples become harder to polarize as coercivity and saturation fields increase with cycling, eventually reaching fields where leakage currents prevent the application of a large enough voltage.

**ACKNOWLEDGMENTS**

The authors are grateful to S. Poissonnet for performing electron microprobe chemical analysis and also R. Guillot for X-Ray diffraction measurements on single crystals. D. Colson wants to acknowledge the Conseil Régional de l'Ile de France (Sésame 2002-2006) and the MIIAT project "Matériaux à Propriétés Remarquables" for their financial supports. This research is supported by the Agence Nationale de la Recherche, project "FEMMES" NT05-1_45147.

**References**

[1] W. Eerenstein, N.D. Mathur, J.F. Scott, Nature, **442**, 759 (2006)

[2] C. Tabares-Munoz, J.-P. Rivera, A. Bezinges, A. Monnier, H. Schmid, Jpn J. Appl. Phys. **24** supp. 2, 1051 (1985)

[3] T. Zhao, A. Scholl, F. Zavaliche, K. Lee, M. Barry, A. Doran, M.P.Cruz, Y.H. Chu, C. Ederer, N.A. Spaldin, R.R. Das, D.M. Kim, S.H Baek, C.B. Eom, R. Ramesh, Nature Materials **5**, 829 (2006)

[4] J.M. Moreau, C. Michel, R. Gerson, W.J. James, J. Phys. Chem. Solids **32**, 1315 (1971)

[5] I. Sosnowska, T. Peterlin-Neumaier, E. Steichele, J. Phys. C: Solid State Phys. **15**, 4835 (1982)

[6] R.T. Smith, G.D. Achenbach, R. Gerson, W.J. James, J. Appl. Phys. **39** N°1, 70 (1968)

[7] J.B. Neaton, C. Ederer, U.V. Waghmare, N.A. Spaldin, K.M. Rabe, Phys. Rev. B **71**, 014113 (2005)

[8] J. Wang, J.B. Neaton, H. Zheng, V. Nagarajan, S.B. Ogale, B. Liu, D. Viehland, V. Vaithyanathan, D.G. Schlom, U.V. Waghmare, N.A. Spaldin, K.M. Rabe, M. Wuttig, R. Ramesh, Science **299**, 1719 (2003)

[9] Y. P. Wang, G.L. Yuan, X.Y. Chen, J.-M. Liu, Z.G. Liu, J. Phys. D: Appl. Phys. **39**, 2019 (2006)

[10] A. K. Pradhan, K. zhang, D. Hunter, J.B. Dadson, G.B. Loutts, P. Bhattacharya, R. Katiyar, J. Zhang, D. J. Sellmer, U. N. Roy, Y. Cui, A. Burger, J. Appl. Phys. **97**, 093903 (2005)

[11] J. R. Teague, R. Gerson, W.J. James, Solid State Comm., **8**, 1073 (1970)

[12] W. Eerenstein, F.D. Morrison, J. Dho, M.G. Blamire, J.F. Scott, N. D. Mathur, Science **307**, 1203a (2005)

[13] D. Lebeugle, D. Colson, A. Forget, M. Viret, P. Bonville, J.F. Marucco, S. Fusil, condmat/0706.0404

[14] F. Kubel, H. Schmid, Acta Cryst., B **46**, 698 (1990)

[15] J. Shieh, J. E. Huber, N. A. Fleck, Journal of The European Ceramic Society **26**, 95 (2006)

[16] C.S. Lynch, L. Chen, W. Yang, Z. Suo, R.M. McMeeking, Journal of Intelligent Material Systems and Structures **6** N°2, 191 (1995)

[17] H. Cao, A.G. Evans, Journal of the American Ceramic Society **77** N°7, 1783 (1994)

# Figure 1

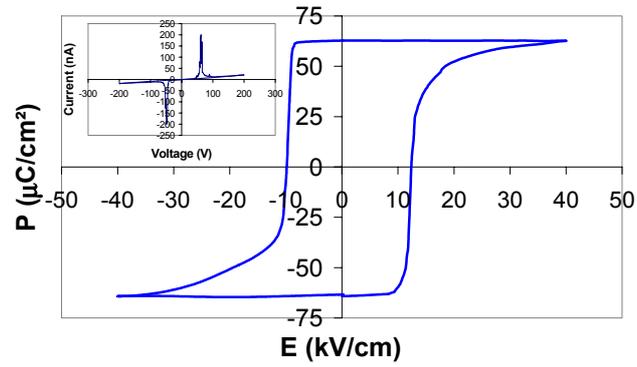

# Figure 2

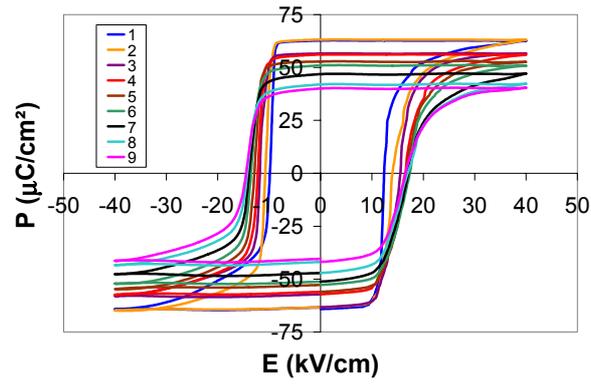

# Figure 3

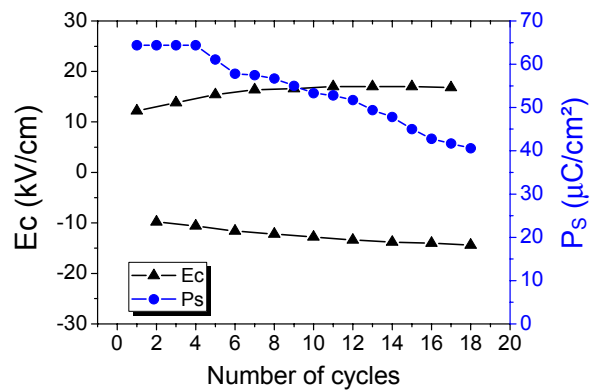

# Figure captions

**Figure 1:** First full P-E hysteresis loop of the single crystal of $BiFeO_3$ at room temperature. The remnant polarization $P_{(012)}$ is 60 $\mu C.cm^{-2}$ and the coercive field is 12 kV/cm. The inferred full saturation polarization along the $[001]_{hex}$ direction is close to 100 $\mu C.cm^{-2}$.
**In insert:** Raw I(V) data.

**Figure 2:** (Color online) Gradual modification of the shape of the P(E) loops during electric field cycling. The loops lose their squareness and the quantities most affected are the coercive and saturation fields as well as the remnant polarization.

**Figure 3:** Decrease of the spontaneous polarization as well as the increase of the coercive field during electric field cycling, as a consequence of a gradual increase of leakage current.